\documentclass[12pt]{iopart}
\usepackage{graphicx}
\usepackage{amssymb}
\usepackage{epsfig}

\newcommand{\ud}{\mathrm{d}}

\begin{document}

\title{Stability of additive-free water-in-oil emulsions.}
\author{Jos Zwanikken, Joost de Graaf, Markus Bier, and Ren\'{e} van Roij,}

\address{Institute for Theoretical Physics, Utrecht
University, Leuvenlaan 4, 3584 CE Utrecht, The Netherlands}

\begin{abstract}
We calculate ion distributions near a planar oil-water interface
within non-linear Poisson-Boltzmann theory, taking into account
the Born self-energy of the ions in the two media. For unequal
self-energies of cations and anions, a spontaneous charge
separation is found such that the water and oil phase become
oppositely charged, in slabs with a typical thickness of the Debye
screening length in the two media. From the analytical solutions,
the corresponding interfacial charge density and the contribution
to the interfacial tension is derived, together with an estimate
for the Yukawa-potential between two spherical water droplets in
oil. The parameter regime is explored where the plasma coupling
parameter exceeds the crystallization threshold, i.e. where the
droplets are expected to form crystalline structures due to a
strong Yukawa repulsion, as recently observed experimentally.
Extensions of the theory that we discuss briefly include numerical
calculations on spherical water droplets in oil, and analytical
calculations of the linear PB-equation for a finite oil-water
interfacial width.
\end{abstract}
\pacs{68.05.-n, 82.70.Kj, 89.75.Fb}

\maketitle

\section{Introduction}
The making and breaking of oil-water emulsions is not only a
problem of extreme importance in chemical, oil, pharmaceutical,
food, and cosmetics industries, but is also a scientifically
fascinating topic. It is well-known that the intrinsic tendency of
oil and water to demix can be slowed-down or delayed by adding
surfactants or colloidal particles to the mixture. These additives
strongly adsorb to the oil-water interface, which lowers the
interfacial tension (the main driving force for demixing) and/or
provides a kinetic barrier that prevents droplet coalescence
\cite{Binks,Sacanna}. Recently however, it was observed that
emulsions of water droplets dispersed in somewhat polar oils can
be stable for a long time (so far for more than 18 months), {\em
without any additives} \cite{Leunissen}. Moreover, the water
droplets, which are of micrometer dimensions in these experiments,
spontaneously form crystalline structures with lattice spacings of
the order of $5-15$ $\mu$m. These observations suggest long-ranged
electrostatic droplet-droplet repulsions due to a net water
droplet charge stemming from a preferential uptake of ions from
the oil \cite{Leunissen,Leunissen2,Zwanikken}. Theoretical
calculations based on Poisson-Boltzmann theory for monovalent ions
in the geometry of a planar water-oil interface showed that the
order of magnitude of the charge separation process, caused by the
different cationic and anionic Born self-energies in oil and
water, is indeed sufficiently strong to explain the observed
stability and crystallization \cite{Leunissen2,Zwanikken}, at
least qualitatively. These predictions are based on the assumption
of a pairwise screened-Coulomb potential of the charged water
droplets through the oil, and an explicit empirical
crystallization condition based on simulations
\cite{Hamaguchi,Vaulina}.

In this paper we will explore the ionic charge separation at the
planar oil-water interface and the resulting crystallization
regime of water droplets dispersed in oil in the high-dimensional
parameter space of salt concentration, dielectric constant of the
oil, ionic sizes and self-energies, droplet size, and droplet
concentration. The analytic solution admitted by the nonlinear
Poisson-Boltzmann equation in the planar geometry
\cite{Verweij,Kung} allows for such a detailed exploration
efficiently; the effects of droplet curvature were recently
studied numerically \cite{deGraaf}. We will find that
crystallization of water droplets is only possible for
sufficiently large droplet radii ($\gtrsim100$ nm), sufficiently
large water content (volume fraction $\gtrsim10^{-3}$), and
sufficiently (but not too) polar oils with dielectric constants
between $4$ and $10$. In addition we will also give an explicit
expression for the change of the interfacial tension between oil
and water due to the ionic segregation over the two media. This
change is found to be negative, proportional to the square-root of
the ionic strength, and small on the scale of the bare oil-water
tension, at least within the present model. We will discuss the
sensitivity of these findings to details of the interfacial
structure as described recently in \cite{Bier}.

\section{Poisson-Boltzmann theory for a planar oil-water interface}
\begin{figure}[!ht]
\centering
\includegraphics[width = 8.1cm]{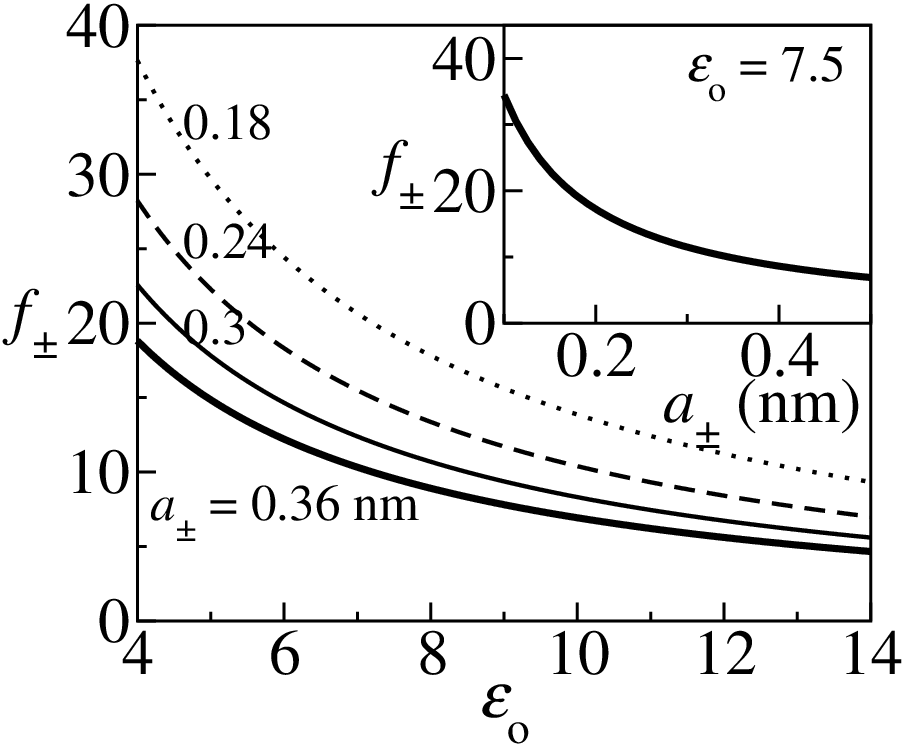}
\caption{The Born self-energy $f_\pm$ in units of $k_\mathrm{B}T$
as a function of dielectric constant $\epsilon_\mathrm{o}$, for
several ionic radii $a_\pm$, and as a function of the ionic radius
at $\epsilon_\mathrm{o}=7.5$ (inset).} \label{fig:Figure1}
\end{figure}

The liquids are considered as structureless homogeneous linear
dielectric media, filling the two half spaces $z<0$ (water) and
$z>0$ (oil), forming a flat interface at $z=0$. The relative
dielectric constant is a step function, $\epsilon(z) =
\epsilon_\mathrm{w}$ $(z<0)$ and $\epsilon(z) =
\epsilon_\mathrm{o}$ $(z>0)$. The grand potential functional per
unit area of the variational density profiles $\rho_\pm(z)$ of the
cations $(+)$ and anions $(-)$ in units of $k_\mathrm{B}T\equiv
1/\beta$ is given by \cite{Zwanikken,deGraaf,Bier,Evans}
\begin{equation}\label{fml:GrandPotentialFunctional}
\beta \Omega [\rho_\pm] = \sum \limits_{\alpha = \pm} \int
\limits_{-\infty}^{\infty} \ud z\ \rho_\alpha(z)
\Big(\ln\frac{\rho_\alpha(z)}{\rho_\mathrm{w}} - 1 +\beta
V_\alpha(z) + \frac{\alpha}{2} \phi(z,[\rho_\pm])\Big),
\end{equation}
with the self-consistent dimensionless electrostatic potential
$\phi(z,[\rho_\pm])$, and with the external potential acting on
the ions
\begin{equation}
\displaystyle\beta V_\pm(z) = \left\{
\begin{array}{ll}
\displaystyle 0&, z<0; \nonumber\\
\displaystyle  \frac{e^2}{2 a_\pm
k_\mathrm{B}T}(\frac{1}{\epsilon_\mathrm{o}}-\frac{1}{\epsilon_\mathrm{w}})
+ g_\pm \equiv f_\pm&, z>0,
\end{array}
\right.
\end{equation}
representing the Born self-energy and an additional specific
solvation energy $g_\pm$ (e.g. due to hydration, hydrogen bonding,
local modification of dielectric constant
\cite{Luo,Onuki,Onuki2}), which is set either to $0$ or to 4 here
in order to study specific effects. The salt concentration in bulk
water is denoted by $\rho_\mathrm{w}$, $e$ is the elementary
charge, and $a_\pm$ is the ionic radius. Both $\phi(z)$ and
$V_\pm(z)$ are gauged to zero in the bulk water phase
($z\rightarrow-\infty$). Figure \ref{fig:Figure1} shows $f_\pm$ as
a function of the oil dielectric constant $\epsilon_\mathrm{o}$,
and ionic radius $a_\pm$ (inset), for $g_\pm=0$. Typical values
for the self-energies are $5-20$ $k_\mathrm{B}T$, and self-energy
differences are of the order of $1-10$ $k_\mathrm{B}T$ for ionic
radii differing by, say, 1 \AA.

\noindent Minimization of the functional
(\ref{fml:GrandPotentialFunctional}) with respect to $\rho_\pm$,
together with the Poisson equation, yields
\begin{equation}\label{fml:Poisson}
\rho_\pm(z) = \rho_\mathrm{w} \exp(-\beta V_\pm(z)\mp\phi(z)),
\end{equation}
which can be rewritten as the Poisson-Boltzmann equation
\begin{equation}\label{fml:PB}
\phi''(z)= \left\{
\begin{array}{ll}
\kappa_\mathrm{w}^2 \sinh(\phi(z)) &, z<0; \nonumber\\
\kappa_\mathrm{o}^2 \sinh(\phi(z)-\phi_\mathrm{D})&, z>0,
\end{array} \right.
\end{equation}
subject to the Neumann boundary conditions
\begin{eqnarray}\label{fml:BoundaryConditions}
\lim \limits_{z\rightarrow-\infty} \phi'(z) &=& \lim
\limits_{z\rightarrow\infty}
\phi'(z)=0; \nonumber \\
\lim \limits_{z\uparrow0} \epsilon_\mathrm{w} \phi'(z) &=&  \lim
\limits_{z\downarrow0} \epsilon_\mathrm{o} \phi'(z),
\end{eqnarray}
which are dictated by the vanishing of the electric displacement
field in the bulk and its continuity at the interface. The Donnan
potential of the bulk oil is $\phi(z\rightarrow\infty)\equiv
\phi_\mathrm{D} = \frac12 ( f_- - f_+)$, the inverse Debye
screening lengths are given by $\kappa_\mathrm{i} = \sqrt{8\pi e^2
\rho_\mathrm{i}/\epsilon_\mathrm{i} k_\mathrm{B}T}$,
$\mathrm{i}\in\{\mathrm{o},\mathrm{w}\}$, and the bulk salt
concentration in oil satisfies $\rho_\mathrm{o} = \rho_\mathrm{w}
\exp(-\frac12 (f_- + f_+))$. The general solution of
(\ref{fml:PB}) is
\begin{equation}\label{fml:ExactSolPB}
\phi(z) = \left\{
\begin{array}{ll}
4\ \mathrm{arctanh}(C_\mathrm{w} e^{ \kappa_\mathrm{w} z})&, z<0;\\
4\ \mathrm{arctanh}(C_\mathrm{o} e^{- \kappa_\mathrm{o} z}) +
\phi_\mathrm{D}&, z>0,
\end{array} \right.
\end{equation}
with integration constants that follow from the boundary
conditions (\ref{fml:BoundaryConditions}) as
\begin{eqnarray}\label{fml:Cow}
C _\mathrm{w} &=& \frac{ n + \cosh\frac{\phi_\mathrm{D}}{2} - D }{
\sinh\frac{\phi_\mathrm{D}}{2} },\nonumber\\
C _\mathrm{o} &=& -\frac{1 +  n\cosh\frac{\phi_\mathrm{D}}{2} - D
}{ n \sinh\frac{\phi_\mathrm{D}}{2}  },
\end{eqnarray}
where $n \equiv \epsilon_\mathrm{w}
\kappa_\mathrm{w}/\epsilon_\mathrm{o} \kappa_\mathrm{o}$, and
$D\equiv \sqrt{ n^2 + 2n \cosh\frac{\phi_\mathrm{D}}{2} + 1}$. The
derivation so far is equivalently presented in
\cite{Verweij,Kung}. A separation of charge is found near the
interface for unequal self-energies, comprising a cloud of net
charge in the water phase of typical width
$\kappa_\mathrm{w}^{-1}$ and one in the oil phase of typical width
$\kappa_\mathrm{o}^{-1}$, with a charge per area at the water side
given in units of $e$ by
\begin{equation}\label{fml:SigmaW}
\sigma_\mathrm{w} = \int \limits_{-\infty}^{0} \ud z
(\rho_+(z)-\rho_-(z))=-\frac{8\rho_\mathrm{w}}{\kappa_\mathrm{w}}\frac{C_\mathrm{w}}{1-C_\mathrm{w}^2};
\end{equation}
the compensating charge resides at the oil side as the system is
globally neutral. In section \ref{sec:Yukawa} we will assume that
the surface charge density of a water droplet in oil is equal to
$\sigma_\mathrm{w}$ in order to calculate the pair potential
between the droplets.

\section{Screened Coulomb potential between water droplets in
oil}\label{sec:Yukawa} We now consider $N$ oil-dispersed water
droplets of radius $a$ in a volume $V$, such that the typical
droplet-droplet separation is $R=(V/N)^\frac{1}{3}$ and the water
volume fraction is $x=\frac43\pi a^3 N/V=\frac43\pi(a/R)^3$. We
assume that each water droplet has a charge $Z=4\pi a^2
\sigma_\mathrm{w}$, where $\sigma_\mathrm{w}$ follows from the
nonlinear PB theory (\ref{fml:SigmaW}), and reconsider a screened
Coulomb (Yukawa) potential between two droplets
\begin{equation}\label{fml:Yukawa}
V_\mathrm{Yuk}(r)= \frac{ (Ze)^2 }{ \epsilon_\mathrm{o} }\Big(
\frac{ e^{\kappa_\mathrm{o} a} }{ 1 + \kappa_\mathrm{o} a }
\Big{)}^2 \frac{ e^{-\kappa_\mathrm{o} r} }{ r },
\end{equation}
with $r$ the center-to-center separation between the droplets.
According to simulation results, crystallization of such a Yukawa
system occurs if $\Gamma\gtrsim 106$ \cite{Hamaguchi,Vaulina},
with the coupling parameter $\Gamma$ defined by
\begin{equation}
\label{fml:CouplingConst} \Gamma \equiv \beta V_\mathrm{Yuk}(R) (1
+ \kappa_\mathrm{o}R + \frac{(\kappa_\mathrm{o}R)^2}2 ).
\end{equation}
Although this condition has been confirmed only for point-Yukawa
systems, it is expected to hold also for finite droplets as long
as $V_\mathrm{Yuk}(2a)\gtrsim 10$ $k_\mathrm{B}T$ and
$\kappa_\mathrm{o} a \lesssim 1$, conditions that are easily met
for micron-sized droplets, provided $4<\epsilon_\mathrm{o}<10$ as
can be seen in figure \ref{fig:Figure3}.

The electrostatic contribution $\gamma$ to the interfacial tension
can be calculated analytically by evaluating the functional
(\ref{fml:GrandPotentialFunctional}) with the equilibrium
profiles, minus the functional evaluated with the step profile
$\rho_\pm(z<0)=\rho_\mathrm{w}$, $\rho_\pm(z>0)=\rho_\mathrm{o}$
\cite{deGraaf,Bier,Onuki2}. After a tedious but straightforward
calculation we find
\begin{equation}
\beta \gamma = -\frac{16\rho_\mathrm{w}}{\kappa_\mathrm{w}}
\frac{C_\mathrm{w}}{1-C_\mathrm{w}^2}(C_\mathrm{w}-C_\mathrm{o})\propto
-\sqrt{\rho_\mathrm{w}} ,
\end{equation}
which is symmetric under exchanging
$\mathrm{w}\leftrightarrow\mathrm{o}$ and where the
proportionality to $\sqrt{\rho_\mathrm{w}}$ follows from
$\kappa_\mathrm{w}\propto\sqrt{\rho_\mathrm{w}}$ and the fact that
$C_\mathrm{w}$ and $C_\mathrm{o}$ are independent of
$\rho_\mathrm{w}$. Typical values of the excess interfacial
tension are at most $\mathcal{O}$($\mu$N/m), and are therefore
vanishingly small compared to the value of the bare oil-water
interfacial tension ($\mathcal{O}$(10 mN/m)).

\section{Numerical results}
\begin{figure}[!ht]
\centering
\includegraphics[width = 7.9cm]{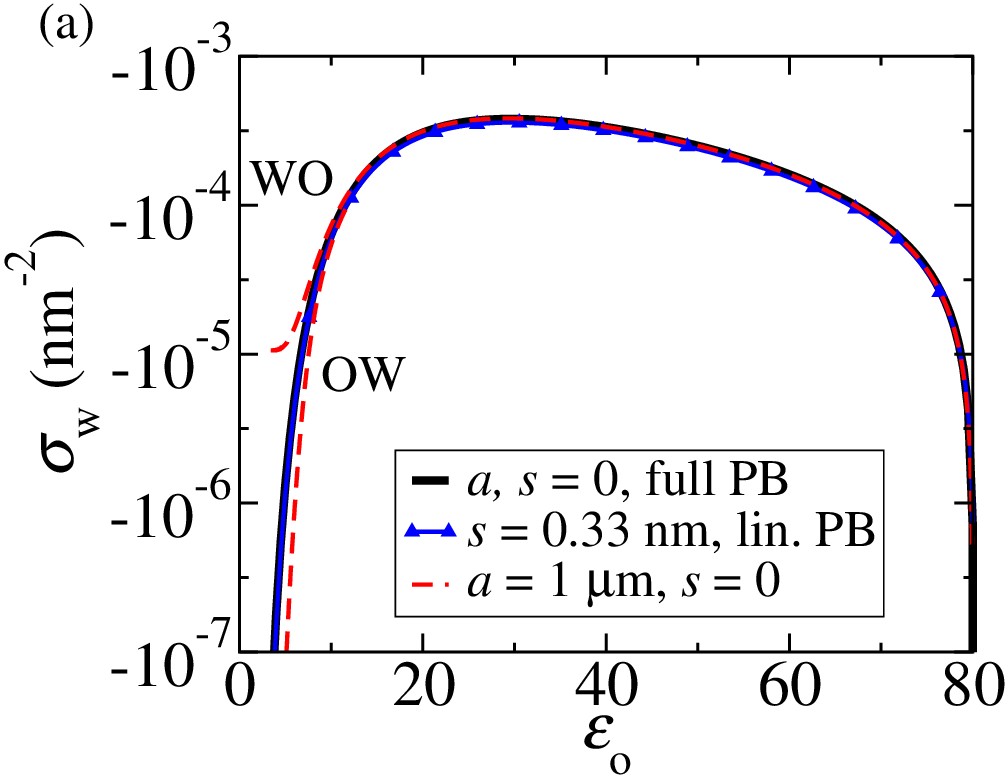}
\hspace{0.5 cm}
\includegraphics[width = 7cm]{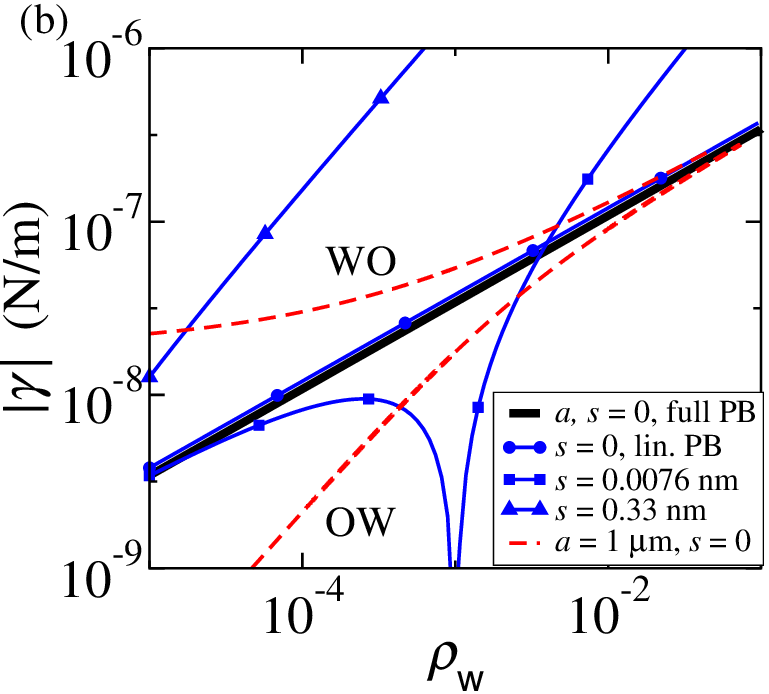}
\caption{(a) Interfacial charge density as a function of the oil
dielectric constant. (b) Excess interfacial tension as a function
of ionic strength, at $\epsilon_\mathrm{o}=7.5$. If the
interfacial width $s=0$, the planar calculations clearly show a
square root behaviour. The sign of $\gamma$ is negative. The
dashed lines show the numerical results from the PB-theory in
spherical geometry, for water droplets of radius $a=1$ $\mu$m
(WO), and equal sized oil droplets in water (OW) \cite{deGraaf}.
The thin lines show the result of the planar linearized PB-theory
with interfacial width $s$, for which $\gamma$ has a linear
asymptote and positive sign in the high-$\rho_\mathrm{w}$ limit
\cite{Bier}, if $s\neq 0$.} \label{fig:Figure2}
\end{figure}

\noindent As standard parameters we use $\epsilon_\mathrm{o}=7.5$,
$\rho_\mathrm{w}=10^{-3}$ M, which are close to the experimental
values of \cite{Leunissen,Leunissen2}, $g_\pm=0$ and
$(a_+,a_-)=(0.36,0.3)$ nm. The surface charge density
$\sigma_\mathrm{w}$ of the interface appears to be strongly
dependent on the oil dielectric constant, keeping the ionic radii
fixed. Typically it is of the order of $\mathcal{O}(10^{-4})$
elementary charges per nm$^{2}$ in the range
$10<\epsilon_\mathrm{o}<70$, and decays rapidly to $10^{-7}$
nm$^{-2}$ for $\epsilon_\mathrm{o}=4$, see figure
\ref{fig:Figure2}(a). Numerical calculations in the spherical
geometry \cite{deGraaf} predict a smaller $\sigma_\mathrm{w}$ for
oil-in-water droplets, and a larger one for water-in-oil droplets
in the regime $4<\epsilon_\mathrm{o}<10$, as shown in figure
\ref{fig:Figure2}(a) for a radius $a=1$ $\mu$m. The excess
interfacial tension $\gamma$ differs correspondingly for finite
droplets, figure \ref{fig:Figure2}(b). The linearized PB theory
that takes into account a finite interfacial width $s\neq 0$
\cite{Bier} between oil and water agrees quantitatively with the
present results for $\sigma_\mathrm{w}$, indicating that the
charge separation is hardly dependent on $s=\mathcal{O}$($10^{-1}$
nm), even if the ions are effectively excluded in a band of
several tenths of nanometers. On the other hand, above a certain
crossover ionic strength, the same theory predicts a qualitatively
different asymptotic behaviour of the excess interfacial tension,
being positive and proportional to the ionic strength
$\rho_\mathrm{w}$, figure \ref{fig:Figure2}(b). One can understand
that as follows: whereas the charge separation is hardly affected
by the interfacial width $s$, the adsorption of particles, and
hence the interfacial tension, is sensitive to the effective
exclusion in a region of width $s$. As the absolute value of the
excess interfacial tension is vanishingly small compared to the
bare oil-water interface, $\gamma_\mathrm{ow}=\mathcal{O}(10$
mN/m), $\gamma$ will be considered not to contribute to the
stability of the system of interest here.

\begin{figure}[!ht]
\centering
\includegraphics[width = 8cm]{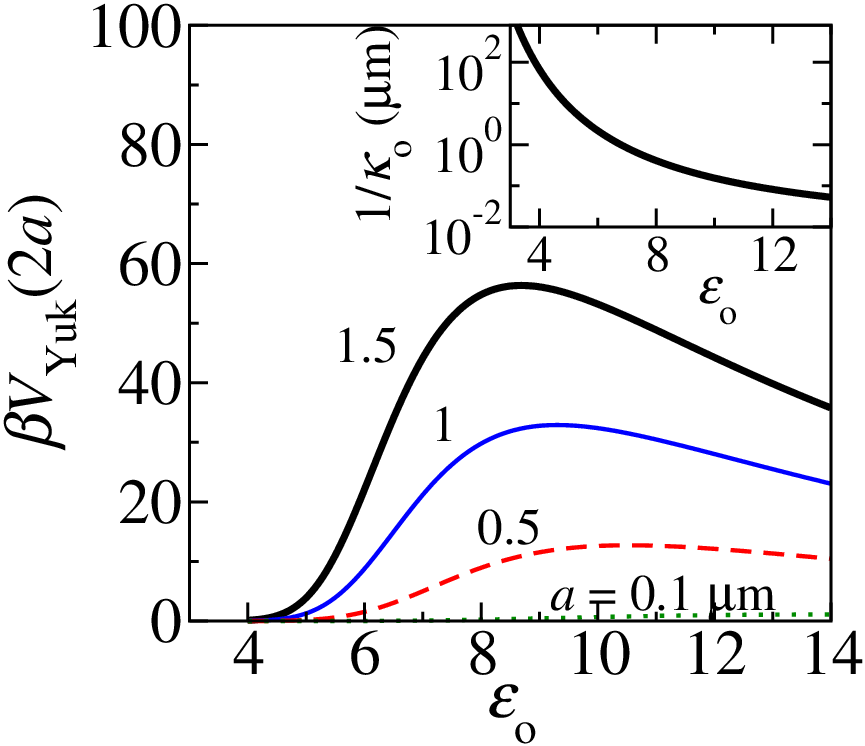}
\caption{The Yukawa potential at contact and the Debye screening
length in oil (inset) as a function of the oil dielectric constant
$\epsilon_\mathrm{o}$, at $\rho_\mathrm{w}=10^{-3}$ M,
$(a_+,a_-)=(0.36,0,3)$. The contact value $V_\mathrm{Yuk}(2a)>1$ $
k_\mathrm{B}T$ only for $\epsilon_\mathrm{o}>4$ and droplet radii
$a>100$ nm. On the other hand, the range of the Yukawa potential
decays rapidly for $\epsilon_\mathrm{o}>10$ (inset); for droplets
with a radius $a>100$ nm, the value of $\kappa_\mathrm{o} a \gg 1$
above $\epsilon_\mathrm{o}\gtrsim 12$. Therefore, significant long
range Coulomb interactions are only expected between $4\lesssim
\epsilon_\mathrm{o} \lesssim 10$. } \label{fig:Figure3}
\end{figure}

\noindent We expect that water droplets in oil can only form
crystalline structures if long range interactions are present
($\kappa_\mathrm{o}a\lesssim 10$), and if the repulsion is
sufficiently strong. Figure \ref{fig:Figure3} shows the Yukawa
potential (\ref{fml:Yukawa}) at contact, $V_\mathrm{Yuk}(2a)$, for
several droplet sizes as a function of $\epsilon_\mathrm{o}$,
revealing that these two conditions already impose strong
restrictions on the dielectric constant and size of the particles.
The inset of figure \ref{fig:Figure3} shows for instance that the
screening length in oil decays from $10-100$ $\mu$m at
$\epsilon_\mathrm{o}=4$ to 100 nm at $\epsilon_\mathrm{o}=12$, for
$\rho_\mathrm{w}=1$ mM, while the main part of figure
\ref{fig:Figure3} shows contact potentials that exceed $10$
$k_\mathrm{B}T$ for $\epsilon_\mathrm{o}\simeq 6-12$ provided the
droplet size is in the micron regime. Thus one only expects
crystallization to be possible for micron-sized water droplet in
oils that are sufficiently polar ($\epsilon_\mathrm{o} \gtrsim 4$)
to have enough charge (see also figure \ref{fig:Figure2}(a)), but
not too polar ($\epsilon_\mathrm{o}\lesssim 12$) to have a long
enough range $\kappa_\mathrm{o}^{-1}$ of the repulsions. This
regime of $\epsilon_\mathrm{o}$ is in remarkably good agreement
with the experimentally found regime \cite{Leunissen,Leunissen2}.

\begin{figure}[!ht]
\centering
\includegraphics[width = 6cm]{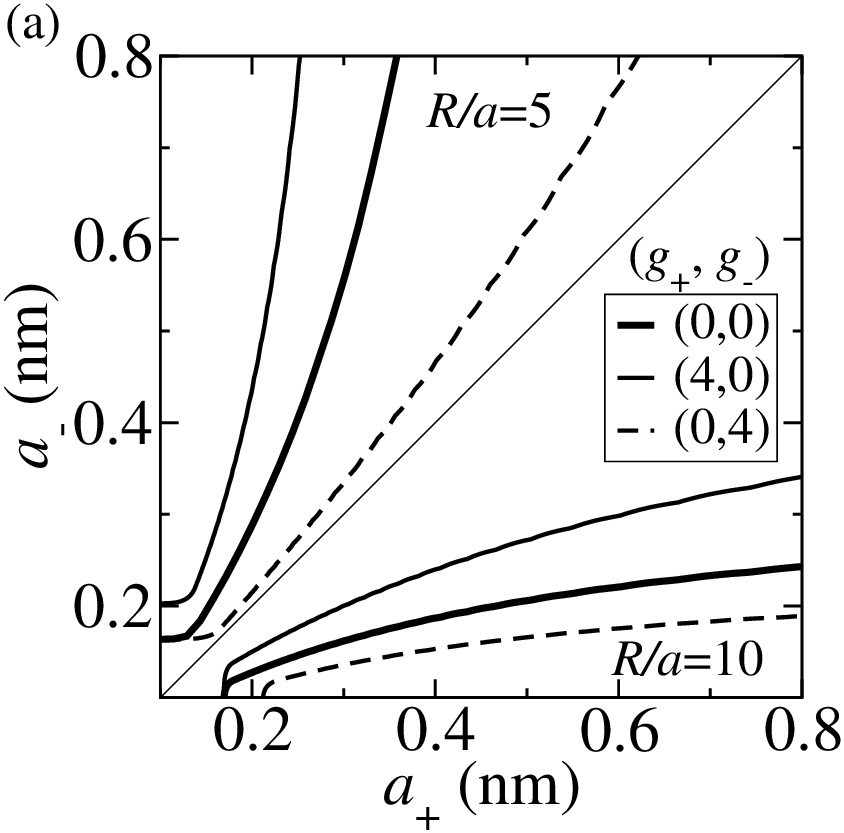}
\hspace{0.5cm}
\includegraphics[width = 6.25cm]{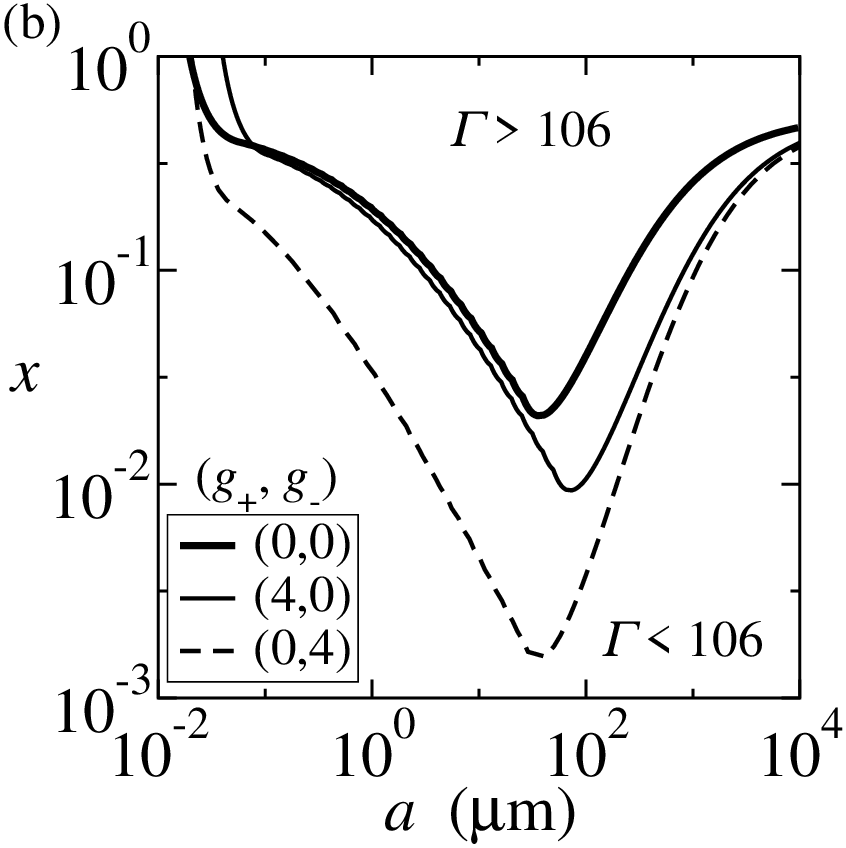}
\caption{(a) The ionic radii $(a_+,a_-)$ for which $\Gamma>106$
for some physically achievable ionic strength $\rho_\mathrm{w}$.
The droplets crystallize in the area between the curves and the
horizontal axis ($R/a=10$), and between the curves and the
vertical axis ($R/a=5$), respectively. The lines can be mirrored
in the diagonal since there is a symmetry under
$a_+\leftrightarrow a_-$, together with $g_+ \leftrightarrow g_-$.
The thin and dashed lines show the results if one of the ionic
species has an additional solvation energy (independent of
$\rho_\mathrm{w}$ and $\epsilon_\mathrm{o}$) of $g_\pm=4$. The
dielectric constant of the oil is $\epsilon_\mathrm{o} = 7.5$ and
droplet radius $a=1.5$ $\mu$m. (b) The volume fractions of water
$x$ and droplet radii $a$ for which $\Gamma>106$ for some
physically achievable ionic strength $\rho_\mathrm{w}$, for
several combinations for $g_\pm$.} \label{fig:Figure4}
\end{figure}

The actual parameter regime where droplets are expected to form
crystalline structures, on the basis of $\Gamma>106$, is depicted
in figure \ref{fig:Figure4}, where the varied parameters are the
ionic radii $a_+,a_-$ (figure \ref{fig:Figure4}(a)) and the volume
fraction and droplet radius $x,a$ (figure \ref{fig:Figure4}(b)).
The lines show the envelopes of the regimes $\Gamma>106$ for all
physically achievable ionic strengths $\rho_\mathrm{w}$.
Variations of the external potential $\beta V_\pm(z)$ by taking a
non-zero $g_\pm$ are seen to result into a significantly modified
crystallization regime. Other solvation effects, e.g. hydration,
hydrogen bonding, local alignment of dipolar fluid molecules, are
therefore expected to be important for a proper quantitative
picture. We examined this for $g_\pm=4$ (figure
\ref{fig:Figure4}), which is considerable compared to
$|f_+-f_-|\simeq 2$ for the present parameters but small compared
to $f_\pm\simeq 12$. Figure \ref{fig:Figure4}(b) shows that
micron-sized droplets, for our standard parameter set, tend to
crystallize at $x \simeq 10^{-2}-10^{-1}$, which is somewhat
higher than the experimentally observed regime $x \simeq
10^{-3}-10^{-2}$ \cite{Leunissen,Leunissen2}. A similarly
relatively high theoretical value for the crystallization volume
fraction was found in \cite{deGraaf}, where curvature effects were
studied, with $g_\pm\equiv0$. Combining these results with those
of figure \ref{fig:Figure4}(b) suggests that the experimental
results may only be quantitatively explained by taking both
curvature and specific solvation effects ($g_\pm\neq 0$) into
account. This problem is left for future studies.

\section{Concluding remarks}

By means of a non-linear Poisson-Boltzmann theory, that takes into
account the Born self-energies of the ions \cite{Verweij,Kung},
and an specific solvation energy \cite{Luo,Onuki,Onuki2}, ionic
equilibrium distributions near an oil-water interface were
calculated. A difference in self-energy was found to lead to a
Donnan potential and charge separation near the interface. The
analytical expression for the charge of the medium per area of
interface agrees roughly with numerical results of the
Poisson-Boltzmann theory in spherical geometry \cite{deGraaf},
although the planar charge density is somewhat underestimated
compared to the value for micron-sized water droplets in oils with
$\epsilon_\mathrm{o}<10$. The charge is in good quantitative
agreement with a linearized Poisson-Boltzmann theory that takes
into account a finite interfacial width $s$ \cite{Bier}, showing
independence of the charge on $s$. For slightly polar oils, i.e.
$4<\epsilon_\mathrm{o}<10$, the typical charge of micron-sized
water droplets ranges between $10-1000$ elementary charges and the
Debye screening lengths range between $0.1 < \kappa_\mathrm{o}a <
10$, such that the particles repel each other electrostatically by
many $k_\mathrm{B}T$ even if their centers are several radii
apart. On the basis of an empirical crystallization condition
\cite{Vaulina} we found a relatively narrow regime in the high
dimensional parameter space, where water droplets in oil are
expected to form crystalline structures, namely only when
$4<\epsilon_\mathrm{o}<10$ (in striking agreement with
experimental findings \cite{Leunissen}), $x \gtrsim 0.001$,
$a\gtrsim 100$ nm, and a sufficiently large difference between the
ionic self-energies $|f_+ - f_-|\gtrsim 1$ $k_\mathrm{B}T$. The
spontaneous charging of the water droplets might play an important
role in the production and stabilization of emulsions with rather
polar oils, since we have seen that by a judicious choice of the
type of oil and salt ions, water droplets can be stabilized by
these, sometimes surprisingly strong, electrostatic effects alone.

This work is part of the research program of the 'Stichting voor
Fundamenteel Onderzoek der Materie' (FOM), which is financially
supported by the 'Nederlandse Organisatie voor Wetenschappelijk
Onderzoek' (NWO).

\end{document}